\documentclass[conference]{IEEEtran}
\IEEEoverridecommandlockouts
\usepackage{cite}
\usepackage{amsmath,amssymb,amsfonts}
\usepackage{algorithmic}
\usepackage{graphicx}
\usepackage{textcomp}
\usepackage{xcolor}
\def\BibTeX{{\rm B\kern-.05em{\sc i\kern-.025em b}\kern-.08em
    T\kern-.1667em\lower.7ex\hbox{E}\kern-.125emX}}
\usepackage{flushend}
\usepackage[breaklinks]{hyperref}
\usepackage{xurl}

\begin{document}

\title{Performance Analysis and Industry Deployment of Post-Quantum Cryptography Algorithms
}

\author{
\IEEEauthorblockN{Elif Dicle Demir}
\IEEEauthorblockA{
    \textit{Electrical and Electronics Eng. Dept.} \\  
    \textit{Koç University}\\
    Istanbul, Türkiye \\
    elifdemir21@ku.edu.tr
}
\and
\IEEEauthorblockN{Buse Bilgin}
\IEEEauthorblockA{
    \textit{6GEN Lab., Next-Gen R\&D} \\  
    \textit{Network Technologies, Turkcell}\\
    Istanbul, Türkiye \\
    buse.bilgin@turkcell.com.tr
}
\and
\IEEEauthorblockN{Mehmet Cengiz Onbaşlı}
\IEEEauthorblockA{
    \textit{Electrical and Electronics Eng. Dept.} \\  
    \textit{Koç University}\\
    Istanbul, Türkiye \\
    monbasli@ku.edu.tr
}
}

\maketitle

\begin{abstract}

As quantum computing advances, modern cryptographic standards face an existential threat, necessitating a transition to post-quantum cryptography (PQC). The National Institute of Standards and Technology (NIST) has selected CRYSTALS-Kyber and CRYSTALS-Dilithium as standardized PQC algorithms for secure key exchange and digital signatures, respectively. This study conducts a comprehensive performance analysis of these algorithms by benchmarking execution times across cryptographic operations such as key generation, encapsulation, decapsulation, signing, and verification. Additionally, the impact of AVX2 optimizations is evaluated to assess hardware acceleration benefits. Our findings demonstrate that Kyber and Dilithium achieve efficient execution times, outperforming classical cryptographic schemes such as RSA and ECDSA at equivalent security levels. Beyond technical performance, the real-world deployment of PQC introduces challenges in telecommunications networks, where large-scale infrastructure upgrades, interoperability with legacy systems, and regulatory constraints must be addressed. This paper examines the feasibility of PQC adoption in telecom environments, highlighting key transition challenges, security risks, and implementation strategies. Through industry case studies, we illustrate how telecom operators are integrating PQC into 5G authentication, subscriber identity protection, and secure communications. Our analysis provides insights into the computational trade-offs, deployment considerations, and standardization efforts shaping the future of quantum-safe cryptographic infrastructure.

\end{abstract}


\section{Introduction}
Modern cryptographic systems rely on the computational intractability of certain mathematical problems, such as integer factorization and discrete logarithms, to ensure the security of digital communication and data protection \cite{joseph2022transitioning}. The advent of quantum computing poses a fundamental threat to modern cryptographic systems, as algorithms such as Shor’s and Grover’s exploit quantum parallelism to break widely used cryptographic primitives. Shor’s algorithm efficiently factors large integers and solves the discrete logarithm problem, undermining the security of RSA and Elliptic Curve Cryptography(ECC), while Grover’s algorithm accelerates brute-force attacks, significantly reducing the effective security of symmetric encryption schemes\cite{bernstein2017post}. As research continues to refine quantum hardware, the urgency to transition towards quantum-resistant cryptographic solutions has become a pressing concern.

To address these emerging threats, the National Institute of Standards and Technology (NIST) initiated the Post-Quantum Cryptography (PQC) Standardization process to develop cryptographic algorithms resilient to quantum threats. The evaluation criteria for candidate algorithms include security against both classical and quantum attacks, cost and performance efficiency, and implementation characteristics such as flexibility and resistance to side-channel attacks\cite{alagic2019status}. As a result of the NIST standardization process, CRYSTALS-Kyber and HQC were selected as key encapsulation mechanisms (KEMs), while CRYSTALS-Dilithium, Falcon, and SPHINCS+ were chosen as digital signature schemes due to their strong security foundations, computational efficiency, and real-world applicability. Kyber is a lattice-based KEM, while HQC is a code-based KEM, both ensuring secure key exchange over insecure communication channels. Similarly, Dilithium and Falcon are lattice-based digital signature schemes designed for message authenticity and integrity, whereas SPHINCS+ is a hash-based scheme. \cite{nist_pqc}

This study focuses on the performance evaluation of post-quantum cryptographic algorithms, specifically Kyber and Dilithium, by benchmarking their execution times across key cryptographic operations. Given the critical role of computational efficiency in the real-world adoption of PQC, our analysis provides insights into their feasibility for practical deployment. Additionally, as the transition to quantum-safe cryptography involves not only technical performance but also industry-wide adoption challenges, we extend our study to include an industry perspective, assessing the implications of PQC deployment in telecommunications and broader enterprise environments.

\section{Testing Methodology and Environment Setup for Performance Analysis}

To understand their computational feasibility, we conducted a detailed performance analysis of Kyber and Dilithium under controlled benchmarking conditions. The performance of cryptographic algorithms is a critical factor in their real-world adoption, particularly in PQC, where computational efficiency directly impacts practical deployment in constrained environments. This section presents a benchmarking study of Kyber and Dilithium, evaluating their execution time across key operations such as key generation, encapsulation, decapsulation, signing, and verification. Additionally, optimizations leveraging AVX2 vector instructions are examined to assess the impact of hardware acceleration on performance. Furthermore, we compare these PQC algorithms with widely used classical cryptographic schemes—Elliptic Curve Diffie-Hellman (ECDH), Elliptic Curve Digital Signature Algorithm (ECDSA), and RSA—to analyze the trade-offs in execution time and efficiency when transitioning to quantum-resistant cryptography.

Each cryptographic operation was executed 1,000 times to ensure consistency, with median and average execution times recorded. The benchmarking methodology follows standard cryptographic evaluation practices, converting measured CPU cycles to execution time using a fixed 3.3 GHz clock. We evaluated both reference and AVX2-optimized implementations of Kyber and Dilithium to assess the performance gains from vectorized instructions. Additionally, to compare PQC with classical cryptography, we tested ECDH, ECDSA, and RSA under the same conditions using OpenSSL libraries.

\section{Performance Evaluation of Kyber and Dilithium}

Table~\ref{tab:kyber_performance} presents performance metrics for Kyber, a key encapsulation mechanism (KEM). It includes the secret key (sk), public key (pk), and ciphertext (ct) sizes for different security levels, reflecting storage and transmission overhead. The listed cryptographic operations are key generation (gen), responsible for producing the key pair; encapsulation (enc), encrypting a shared secret using the recipient’s public key; and decapsulation (dec), recovering the shared secret with the private key. Table ~\ref{tab:dilithium_performance} provides results for Dilithium, a digital signature scheme. It reports public key (pk) and signature (sig) sizes, which indicate storage costs for authentication. The benchmarked operations include key generation (gen), used to create the signing key pair; signing (sign), which generates digital signatures for message integrity; and verification (verify), ensuring the validity of signatures. The AVX2 speedup rate in Tables I and II represents the performance improvement of the AVX2-optimized implementation compared to the reference implementation. It is calculated as the ratio of execution times, indicating how many times faster the AVX2 implementation performs a given cryptographic operation. A higher speedup value signifies greater efficiency gains achieved through vectorized polynomial arithmetic in AVX2-enabled processors.

As indicated in Table~\ref{tab:kyber_performance}, the execution times of Kyber increase with higher security levels across all three operations: key generation, encapsulation, and decapsulation. Notably, Kyber-512 completes execution in 0.127 ms, whereas Kyber-1024 requires 0.294 ms, demonstrating the expected computational cost of increased cryptographic strength. However, the scaling is nonlinear, as the increase from Kyber-768 to Kyber-1024 is smaller than from Kyber-512 to Kyber-768.

The AVX2 optimization significantly reduces execution time, yielding an average speedup of 5.98× across different security levels. The most substantial gains occur in decapsulation, which is reduced by up to 6.65× due to the vectorized polynomial arithmetic enabled by AVX2 instructions. This demonstrates that Kyber benefits greatly from parallelization, making it well-suited for optimized hardware implementations.

Similarly, as shown in Table~\ref{tab:dilithium_performance}, the execution time of Dilithium scales with security levels, with Dilithium-2 executing in 0.643 ms while Dilithium-5 requires 1.36 ms. Unlike Kyber, where operations are relatively balanced, Dilithium’s signing step dominates execution time—accounting for over 60\% of the total runtime in all security levels. This is due to the structured lattice sampling required for signature generation, which is inherently more computationally expensive than verification.

The AVX2 speedup for Dilithium is lower than for Kyber (4.8× on average), but still significant, particularly in the signing operation, which achieves up to a 5.83× reduction in execution time. The verification step sees the smallest speedup (3.76×), reflecting its already efficient nature. The results emphasize that while Dilithium is computationally heavier than Kyber, its AVX2-optimized variant brings notable efficiency improvements, making it feasible for real-world applications.

Overall, the results in Tables~\ref{tab:kyber_performance} and \ref{tab:dilithium_performance} underscore the computational viability of Kyber and Dilithium, demonstrating that hardware optimizations (e.g., AVX2) significantly enhance performance. These findings highlight the practicality of post-quantum cryptography (PQC) deployment, as even without specialized hardware accelerators, Kyber and Dilithium achieve efficient execution times while maintaining high security.

\begin{table}[h]
\centering
\caption{Key and ciphertext sizes and execution times (in milliseconds) for all parameter sets of Kyber.}
\renewcommand{\arraystretch}{1.2}
\begin{tabular}{|c|c|c|c|}
\hline
\multicolumn{4}{|c|}{\textbf{KYBER 512}} \\
\hline
\textbf{Sizes (Bytes)} & \textbf{Reference (ms)} & \textbf{AVX2 (ms)} & \textbf{AVX2 Speedup Rate} \\
\hline
sk: 1632 & gen: 0.035 & gen: 0.007 & 5.00 \\
pk: 800  & enc: 0.040 & enc: 0.007 & 5.71 \\
ct: 768  & dec: 0.052 & dec: 0.008 & 6.50 \\
\hline
\textbf{Total} & 0.127 & 0.022 & 5.77 \\
\hline
\multicolumn{4}{|c|}{\textbf{KYBER 768}} \\
\hline
\textbf{Sizes (Bytes)} & \textbf{Reference (ms)} & \textbf{AVX2 (ms)} & \textbf{AVX2 Speedup Rate} \\
\hline
sk: 2400 & gen: 0.058 & gen: 0.011 & 5.27 \\
pk: 1184 & enc: 0.063 & enc: 0.011 & 5.73 \\
ct: 1088 & dec: 0.080 & dec: 0.012 & 6.67 \\
\hline
\textbf{Total} & 0.201 & 0.034 & 5.91 \\
\hline
\multicolumn{4}{|c|}{\textbf{KYBER 1024}} \\
\hline
\textbf{Sizes (Bytes)} & \textbf{Reference (ms)} & \textbf{AVX2 (ms)} & \textbf{AVX2 Speedup Rate} \\
\hline
sk: 3168 & gen: 0.089 & gen: 0.015 & 5.93 \\
pk: 1568 & enc: 0.092 & enc: 0.015 & 6.13 \\
ct: 1568 & dec: 0.113 & dec: 0.017 & 6.65 \\
\hline
\textbf{Total} & 0.294 & 0.047 & 6.26 \\
\hline
\end{tabular}
\label{tab:kyber_performance}
\end{table}

\begin{table}[h]
\centering
\caption{Public key and signature sizes and execution times (in milliseconds) for all parameter sets of Dilithium.}
\renewcommand{\arraystretch}{1.2}
\begin{tabular}{|c|c|c|c|}
\hline
\multicolumn{4}{|c|}{\textbf{DILITHIUM 2}} \\
\hline
\textbf{Sizes (Bytes)} & \textbf{Reference (ms)} & \textbf{AVX2 (ms)} & \textbf{AVX2 Speedup Rate} \\
\hline
pk: 1312 & gen: 0.094 & gen: 0.026 & 3.62 \\
sig: 2420 & sign: 0.445 & sign: 0.077 & 5.78 \\
 & verify: 0.104 & verify: 0.028 & 3.71 \\
\hline
\textbf{Total} & 0.643 & 0.131 & 4.91 \\
\hline
\multicolumn{4}{|c|}{\textbf{DILITHIUM 3}} \\
\hline
\textbf{Sizes (Bytes)} & \textbf{Reference (ms)} & \textbf{AVX2 (ms)} & \textbf{AVX2 Speedup Rate} \\
\hline
pk: 1952 & gen: 0.167 & gen: 0.045 & 3.71 \\
sig: 3293 & sign: 0.665 & sign: 0.120 & 5.54 \\
 & verify: 0.160 & verify: 0.045 & 3.56 \\
\hline
\textbf{Total} & 0.992 & 0.210 & 4.73 \\
\hline
\multicolumn{4}{|c|}{\textbf{DILITHIUM 5}} \\
\hline
\textbf{Sizes (Bytes)} & \textbf{Reference (ms)} & \textbf{AVX2 (ms)} & \textbf{AVX2 Speedup Rate} \\
\hline
pk: 2592 & gen: 0.253 & gen: 0.070 & 3.61 \\
sig: 4595 & sign: 0.840 & sign: 0.144 & 5.83 \\
 & verify: 0.267 & verify: 0.071 & 3.76 \\
\hline
\textbf{Total} & 1.360 & 0.285 & 4.77 \\
\hline
\end{tabular}
\label{tab:dilithium_performance}
\end{table}

\section{Performance Comparison: Post-Quantum Cryptography vs. Classical Cryptography}

Table~\ref{tab:pqc_vs_classical} presents a comparative analysis of execution times for post-quantum and classical cryptographic algorithms, evaluated under controlled conditions. Both PQC and classical schemes were tested at different security levels, measured in bits, to assess performance variations. The table includes cryptographic algorithms across multiple security configurations, ensuring a direct comparison of execution times. The evaluation focuses on total execution time, measured in milliseconds, to quantify computational cost across different cryptographic operations. While this analysis highlights execution speed, real-world deployment must also consider additional factors such as memory footprint, communication overhead, and hardware compatibility.

Kyber exhibits notable performance advantages over both RSA and ECDH, which are commonly employed for key exchange. Kyber-512, offering 128-bit security, achieves execution times that are approximately three times faster than both RSA-2048 and ECDH(P-256), despite these classical schemes providing lower security guarantees. Even Kyber-1024, the most computationally expensive variant, maintains an execution time that is roughly three times faster than RSA-3072, which offers only 128-bit security. At equivalent security levels, Kyber consistently achieves faster execution times than ECDH(P-256, P-384, P-521) while also providing quantum resistance. These efficiency gains are attributed to Kyber’s lattice-based cryptographic foundation, which relies on small polynomials and number-theoretic transforms (NTT) rather than large-number modular exponentiation. This mathematical structure enables faster key generation and encapsulation while maintaining strong security guarantees, particularly against quantum adversaries.

Dilithium demonstrates significant computational advantages over ECDSA, a widely used classical digital signature scheme. At the 128-bit security level, Dilithium-2 executes signature operations approximately 20\% faster than ECDSA(P-256), with the performance gap increasing at higher security levels. Dilithium-5, the highest-security variant, achieves nearly twice the execution speed of ECDSA(P-512) at the 256-bit security level. A distinct characteristic of Dilithium is that signature generation dominates execution time, accounting for over 60\% of the total runtime, whereas ECDSA exhibits a more balanced distribution between signing and verification. This difference arises from Dilithium’s structured lattice sampling, which, while computationally intensive, remains more efficient than ECDSA’s elliptic curve discrete logarithm operations. Additionally, Dilithium’s deterministic signature generation eliminates nonce-related vulnerabilities, a known weakness in ECDSA implementations.

The results indicate that post-quantum cryptographic algorithms do not inherently introduce higher computational costs. On the contrary, Kyber and Dilithium frequently outperform classical cryptographic schemes at equivalent security levels. Kyber consistently demonstrates superior efficiency in key exchange operations compared to RSA and ECDH, even at its highest security configuration. Similarly, Dilithium provides a computationally efficient alternative to ECDSA, particularly as security levels increase. While Dilithium’s signing operation remains computationally heavier than verification, it still surpasses ECDSA in signature generation across all tested configurations. These findings highlight the feasibility of transitioning to quantum-resistant cryptographic standards in practical applications, demonstrating that enhanced security can be achieved without compromising computational efficiency.

These performance findings highlight the computational feasibility of Kyber and Dilithium as post-quantum cryptographic solutions, demonstrating that quantum resistance does not necessarily come at the cost of execution efficiency. However, execution time is only one aspect of cryptographic feasibility. While our controlled benchmarking showed that Kyber and Dilithium outperform classical schemes in speed, these results were obtained under optimized and isolated conditions. Real-world deployment involves additional complexities, such as infrastructure constraints, interoperability with existing systems, and operational overhead, which can impact practical performance. Thus, while PQC shows strong computational efficiency, its large-scale adoption in telecom networks requires a broader evaluation, considering scalability, integration challenges, and regulatory compliance.

\begin{table}[t]
\centering
\caption{Execution time comparison of post-quantum and classical cryptographic algorithms.}
\label{tab:pqc_vs_classical}
\begin{tabular}{|l|c|c|}
\hline
\textbf{Algorithm} & \textbf{Security Level} & \textbf{Total Time (ms)} \\ \hline
Kyber-512 & 128-bit & 0.127 \\ \hline
Kyber-768 & 192-bit & 0.201 \\ \hline
Kyber-1024 & 256-bit & 0.294 \\ \hline
Dilithium-2 & 128-bit & 0.643 \\ \hline
Dilithium-3 & 192-bit & 0.992 \\ \hline
Dilithium-5 & 256-bit & 1.360 \\ \hline
ECDSA(P-256) & 128-bit & 0.801 \\ \hline
ECDSA(P-384) & 192-bit & 1.702 \\ \hline
ECDSA(P-512) & 256-bit & 2.398 \\ \hline
RSA-2048 & 112-bit & 0.324 \\ \hline
RSA-3072 & 128-bit & 0.884 \\ \hline
ECDH(P-256) & 128-bit & 0.102 \\ \hline
ECDH(P-384) & 192-bit & 0.299\\ \hline
ECDH(P-521) & 256-bit & 0.903  \\ \hline
\end{tabular}
\end{table}

\section{Post-Quantum Cryptography in Telecommunications: Challenges, Implementations, and Future Outlook}

Implementing PQC in telecommunications networks presents significant challenges. Telecom operators must upgrade complex, large-scale infrastructures that currently rely on classical encryption, all while maintaining service continuity. Key challenges include performance and latency impacts, compatibility with legacy systems, lack of finalized standards, resource and cost constraints, transitional security risks, and vendor readiness issues.
\subsection{Challenges}
\subsubsection{Performance Impact on Existing Infrastructure}
PQC algorithms require more computational resources and larger key sizes than classical cryptography. Many schemes are at least an order of magnitude slower or produce larger keys and ciphertexts than RSA or ECC, straining network devices~\cite{GSMA2024PQC}. The increased size of PQC keys, signatures, and ciphertexts taxes bandwidth and memory-constrained hardware; for instance, an additional 1 KB in a TLS handshake can increase response time by ~1.5\%~\cite{PQCConference2025}. 
Latency-sensitive telecom applications, such as voice and video, may experience performance degradation due to longer cryptographic operations or larger handshake messages. Operators need to evaluate whether servers, routers, and HSMs can support the increased computational load of PQC, as many may require hardware upgrades specifically for PQC adoption. Especially in radio access networks (RANs) and customer devices with limited processing power, PQC’s computational overhead and memory footprint pose a significant deployment challenge.

\subsubsection{Interoperability with Legacy Systems}
During the transition, not all network elements and partner systems will upgrade to PQC at the same time, raising interoperability issues. If one system uses a PQC-based protocol but the communicating peer does not, secure connections cannot be established~\cite{USPQCReport2024}. Many telecom protocols use a “fail secure” approach, meaning a PQC-enabled node could be cut off from legacy nodes that don’t recognize the new algorithms. Due to the interconnected nature of telecom networks, a single non-upgraded component can block migration, creating deployment bottlenecks. A possible solution is hybrid cryptographic modes (combining classical and PQC algorithms), but this adds complexity and requires new protocol standards and careful validation, potentially slowing down the transition. To prevent network partitioning, telecom operators must ensure PQC upgrades happen in sync across critical systems or remain backward-compatible.

\subsubsection{Standardization and Regulatory Concerns}
The telecom industry is highly standardized and regulated, so PQC adoption hinges on mature standards and regulatory guidance. As of 2024, standards bodies like NIST are just publishing the first official PQC algorithm standards~\cite{Taaffe2023}. Until international standards (e.g., 3GPP, IETF, ETSI) incorporate PQC, telcos risk adopting interim solutions that might not be interoperable or compliant long-term. There is also regulatory pressure: governments and industry bodies are already setting timelines and mandates for quantum-safe transitions. For example, the U.S. National Security Agency’s CNSA 2.0 mandates specific PQC algorithms, aiming all national security systems to be quantum-resistant by 2035. However, inconsistent national strategies pose challenges for global carriers, as many countries have only issued high-level guidance to “start planning” with few concrete standards yet. The absence of finalized telecom-specific PQC standards adds uncertainty, requiring operators to closely coordinate with standards organizations to ensure protocols like 5G authentication, IPsec, and TLS integrate PQC effectively.

\subsubsection{Cost and Resource Allocation}
Upgrading a telecom operator’s cryptographic infrastructure to PQC is costly and resource-intensive. Many legacy systems lack the processing power, memory, or bandwidth to support PQC, requiring replacement or retrofitting of equipment such as mobile devices, SIM cards, routers, and base stations. This represents a significant capital expense, with costs extending to PQC-capable HSMs, accelerator cards, software updates, staff training, testing, and parallel system operation during the transition. Smaller operators worry that only large carriers can afford early adoption, but as vendors integrate PQC into products, upgrade costs are expected to decrease. Nonetheless, operators need to allocate substantial resources for cryptographic inventory, upgrade planning, and continuous maintenance to ensure a smooth migration. The cost of inaction could be higher—a quantum-broken network may result in regulatory penalties and customer loss, making early investment crucial.

\subsubsection{Security Risks and Transition Challenges}
Transitioning to PQC raises security concerns, as these new algorithms have not been tested in real-world deployments for decades like RSA/ECC. There are risks of undiscovered weaknesses or implementation flaws, and some PQC candidates have already been found vulnerable to cryptanalysis and side-channel attacks during standardization. Ensuring side-channel resistance is critical—cryptographic operations must not leak secrets through timing, power, or memory access patterns. Additionally, PQC introduces complex key management and new failure modes; for example, some digital signature schemes require tracking one-time keys, complicating network authentication. Early deployments have exposed issues, such as network middleware and firewalls failing due to large key exchange messages. Misconfigurations, like hybrid mode errors or certificate management lapses, could introduce vulnerabilities. To mitigate these risks, telecom operators must conduct extensive testing, use proven implementations, and ensure crypto-agility, allowing algorithm updates when needed.

\subsubsection{Vendor Readiness and Supply Chain Considerations}
Telecommunications relies on a vast network of vendors for hardware, software, and infrastructure, making PQC adoption a supply chain challenge. Many vendors await finalized standards before integrating PQC, and without support in critical components like SIM cards and routers, full migration is impossible. To address this, telecom operators are updating procurement policies, requiring vendors to support NIST-approved PQC algorithms and crypto-agility. Regulatory bodies may also mandate certification, potentially delaying availability. While some vendors are developing PQC-capable products, widespread readiness will take time. Effective supply chain management and early engagement with suppliers are essential to ensure smooth deployment, coordinated upgrades, and interoperability. Ultimately, achieving a quantum-safe telecom network requires industry-wide collaboration and careful planning.

\subsection{Successful Implementations and Initiatives of PQC}
Despite the challenges, there have been several successful implementations and trials of post-quantum cryptography in telecom contexts. Forward-thinking carriers and technology partners around the world have started to integrate PQC into test networks, demonstrating feasibility and gleaning best practices. Below are a few notable examples and case studies highlighting how PQC deployment is being approached in telecommunications:

\subsubsection{SoftBank (Japan) – Hybrid PQC Network Trial}
SoftBank Corp., a major mobile operator in Japan, partnered with SandboxAQ to test PQC algorithms in a live network environment. In 2023 they conducted a hybrid encryption trial, combining classical elliptic-curve cryptography with lattice-based post-quantum algorithms on live network traffic~\cite{SoftBank2022}. The results were encouraging: the hybrid quantum-safe approach was verified to work on existing 4G/5G infrastructure with minimal performance impact. SoftBank reported that lattice-based PQC algorithms (such as those later standardized by NIST) outperformed other quantum-safe alternatives in their tests, providing strong security with only marginal added latenc~\cite{SoftBank2023blog}. By adopting a hybrid approach, SoftBank ensured interoperability with existing systems while enhancing security. Their phased deployment, from lab tests to real-world networks, demonstrated that careful algorithm selection and optimization can mitigate future quantum threats without major performance trade-offs. Collaboration with SandboxAQ helped streamline cryptographic inventory and regulatory compliance. SoftBank continues investing in PQC, positioning early adoption as a competitive advantage in secure telecom infrastructure.

\subsubsection{SK Telecom (South Korea) – PQC in 5G Standalone Network}
Another pioneering effort was led by SK Telecom (SKT) in South Korea, in collaboration with Thales. SKT and Thales carried out a groundbreaking test of post-quantum cryptography in a real 5G standalone network environment~\cite{ThalesSKT2024}. In this pilot, SKT deployed quantum-resistant encryption to secure subscriber identities and network traffic. They tested 5G USIM cards implementing the CRYSTALS-Kyber key encapsulation algorithm, ensuring authentication remains secure against quantum threats. The trial demonstrated seamless interoperability between PQC-protected SIMs and the core network, with encrypted calls proving quantum-safe communication. This deployment, one of the first PQC integrations in 5G, underscores the role of carrier-vendor partnerships and informs ongoing standards development.

\subsubsection{North American Carriers and Initiatives}
U.S. and Canadian telecom operators are preparing for PQC, driven by government directives. AT\&T plans to be “quantum ready” by 2025, with internal pilots testing PQC in VPNs and TLS. While large-scale deployments are pending, mandates for critical infrastructure are pushing adoption.
In Canada and Europe, providers like Verizon, Rogers, Deutsche Telekom, and BT are engaged in research and industry collaborations, focusing on PQC for routing, customer data protection, and inter-carrier security. These efforts emphasize crypto-agility—ensuring networks can transition flexibly as standards evolve. Industry groups, including the GSMA Post-Quantum Telco Network Taskforce and 5G Americas, are developing best practices to guide telecom operators through PQC adoption.

Across successful implementations, key best practices have emerged. Conducting a cryptographic inventory helps identify necessary upgrades, while pilot deployments in less constrained environments allow for manageable PQC integration. A hybrid approach, running PQC alongside classical encryption, has been widely adopted to maintain continuity. Vendor collaboration with SIM card providers, router manufacturers, and software vendors is crucial for early integration. These trials also confirm that early PQC deployment safeguards critical data from future quantum threats. While performance impacts are generally manageable with optimized algorithms, some operators have even improved network efficiency by modernizing legacy systems. Overall, these case studies demonstrate that with careful planning and phased execution, telcos can begin inserting quantum-resistant cryptography into their networks today, gaining experience and confidence for broader rollouts.

\subsection{Future Outlook and Recommendations}

The adoption of post-quantum cryptography (PQC) in telecom networks is shifting from isolated pilots to broader deployments as standards solidify and the quantum threat looms. Telecom operators must act now, as waiting until large-scale quantum computers emerge will be too late. Security organizations stress the urgency of conducting cryptographic inventories to identify where public-key cryptography is used—such as SIM authentication, SSL/TLS links, and PKI certificates—prioritizing critical assets to mitigate “harvest now, decrypt later” risks. Awareness and education are also crucial for leadership and technical teams.

A structured PQC implementation roadmap involves phased deployments, starting with hybrid cryptographic modes alongside classical encryption to maintain compatibility. Initial transitions should focus on non-customer-facing segments, expanding as standards mature and interoperability improves. Operators must align migration plans with regulatory requirements, ensuring compliance with evolving mandates. From 2024 onward, telecom providers are expected to integrate PQC into technology refresh cycles, with PQC becoming a standard in 5G-Advanced and 6G networks by the late 2020s. The goal is to achieve full quantum resistance in critical infrastructure by the early 2030s.

Within the next decade, PQC will likely be as integral to telecom security as TLS and IPsec are today. Once NIST and other bodies finalize standards by 2024–2025, adoption will accelerate, giving early adopters a competitive edge with “quantum-safe” services. Given the uncertainty of quantum computing advancements, proactive preparation is essential. Encouragingly, PQC adoption does not necessarily require hardware replacements—many transitions can be done via software updates, reducing costs. As vendors integrate PQC into products, expenses are expected to decrease further.

By 2030, much of global telecom traffic, particularly sensitive communications, will likely be encrypted using post-quantum or hybrid cryptographic schemes. Collaboration among telecom operators, governments, and the security community will be crucial for interoperability and resilience. With proactive planning and cooperative execution, the telecom industry can secure global communications against quantum threats while maintaining security, efficiency, and compliance.

\section{Conclusion}
The transition to post-quantum cryptography (PQC) is no longer a theoretical consideration but an imminent necessity for securing digital communications against future quantum threats. This study has demonstrated that CRYSTALS-Kyber and CRYSTALS-Dilithium, the NIST-standardized PQC algorithms, not only provide robust quantum resistance but also achieve competitive execution times compared to classical cryptographic schemes. Benchmarking results highlight their computational efficiency, particularly when optimized with AVX2 vectorization.

However large-scale deployment in telecommunications networks introduces critical challenges, including infrastructure upgrades, interoperability concerns, regulatory compliance, and cost constraints. The successful implementation of PQC in telecom environments requires a structured, phased migration strategy, leveraging hybrid cryptographic approaches to maintain compatibility with legacy systems. Early industry trials demonstrate the viability of PQC adoption while emphasizing the importance of vendor collaboration, cryptographic agility, and thorough performance validation.

Looking ahead, PQC is expected to become a fundamental component of telecom security, with adoption accelerating as standards solidify and regulatory mandates take effect. As quantum computing advances remain unpredictable, proactive preparation is essential to mitigate risks associated with delayed migration. Encouragingly, the ongoing integration of PQC into security protocols for 5G and 6G networks, along with continued industry cooperation, ensures that telecom infrastructure remains resilient against emerging cryptographic threats. With careful planning and strategic execution, the transition to quantum-safe cryptography can safeguard telecom networks, ensuring their security and adaptability in the quantum era.

\bibliographystyle{IEEEtran}

\begin{thebibliography}{99}

\bibitem{joseph2022transitioning}
D. Joseph, R. Misoczki, M. Manzano, J. Tricot, F. D. Pinuaga, O. Lacombe, S. Leichenauer, J. Hidary, P. Venables, and R. Hansen, 
"Transitioning organizations to post-quantum cryptography," Nature, vol. 605, no. 7909, pp. 237–243, 2022.

\bibitem{bernstein2017post}
D. J. Bernstein and T. Lange, "Post-quantum cryptography," Nature, vol. 549, no. 7671, pp. 188–194, 2017.

\bibitem{alagic2019status}
G. Alagic, G. Alagic, J. Alperin-Sheriff, D. Apon, D. Cooper, Q. Dang, Y.-K. Liu, C. Miller, D. Moody, R. Peralta et al., 
"Status report on the first round of the NIST post-quantum cryptography standardization process," 2019.

\bibitem{nist_pqc}
National Institute of Standards and Technology, "Post-Quantum Cryptography Standardization," 2024, accessed: 2024-03-17. [Online]. Available: \url{https://csrc.nist.gov/projects/post-quantum-cryptography/selected-algorithms}.

\bibitem{GSMA2024PQC}
GSM Association, "Post Quantum Cryptography – Guidelines for Telecom Use Cases," GSM Association, Technical Report PQ.03, February 2024, accessed: 2024-03-17. [Online]. Available: https://www.gsma.com/newsroom/wp-content/uploads/PQ.03-Post-Quantum-Cryptography-Guidelines-for-Telecom-Use-v1.0.pdf.

\bibitem{PQCConference2025}
PKI Consortium, "Key takeaways of the PQC conference in Austin," January 30, 2025, accessed: 2025-03-17. [Online]. Available: \url{https://pkic.org/2025/01/30/key-takeaways-of-the-pqc-conference-in-austin/}.

\bibitem{USPQCReport2024}
U. Government, "Report on post-quantum cryptography," Government Report, The White House, Washington, D.C., Tech. Rep. REF PQC-Report FINAL Send, July 2024, presented to the Senate Committee on Homeland Security and Governmental Affairs and the House Committee on Oversight and Accountability. [Online]. Available: \url{https://bidenwhitehouse.archives.gov/wp-content/uploads/2024/07/REF_PQC-Report_FINAL_Send.pdf}



\bibitem{Taaffe2023}
J. Taaffe, "Are telcos ready for a quantum leap?" June 2023, accessed: March 17, 2025. [Online]. Available: https://inform.tmforum.org/features-and-opinion/are-telcos-making-a-quantum-leap.

\bibitem{SoftBank2022}
SoftBank Corp. and SandboxAQ, "SoftBank Corp. and SandboxAQ to Jointly Implement Next-Generation Cryptosystem Resilient to Cyber Attacks from Quantum Computers," March 2022, press Release, accessed: March 17, 2025. [Online]. Available: https://www.sandboxaq.com/press/softbank-corp-and-sandbox-aq-to-jointly-implement-next-generation-cryptosystem-resilient-to-cyber-attacks-from-quantum-computers.

\bibitem{SoftBank2023blog}
SoftBank Corp., "SoftBank Corp. and SandboxAQ Jointly Verify Hybrid Mode Quantum-safe Technology," February 2023, blog Post, accessed: March 17, 2025. [Online]. Available: https://www.softbank.jp/en/corp/technology/research/story-event/008/.

\bibitem{ThalesSKT2024}
Thales Group and SK Telecom, "Thales and SK Telecom: Pioneering Quantum-Resistant Cryptography for 5G Networks," 2024, accessed: March 17, 2025. [Online]. Available: https://www.thalesgroup.com/en/markets/digital-identity-and-security/mobile/5G-skt-post-quantum-user-case.

\end{thebibliography}

\end{document}